\documentclass[fleqn,usenatbib]{mnras}

\usepackage{newtxtext,newtxmath}

\usepackage[T1]{fontenc}

\DeclareRobustCommand{\VAN}[3]{#2}
\let\VANthebibliography\thebibliography
\def\thebibliography{\DeclareRobustCommand{\VAN}[3]{##3}\VANthebibliography}

\usepackage{graphicx}	
\usepackage{amsmath}	
\usepackage{siunitx}
\usepackage{subcaption}
\usepackage{threeparttable}
\usepackage{float}
\usepackage{anyfontsize}
\usepackage{lscape} 
\usepackage{caption}

\title{Discovery of double BSS sequences in the old Galactic open cluster Berkeley 17}

\author[Khushboo K. Rao et al.]{Khushboo K. Rao,$^{1}$\thanks{E-mail: p20170419@pilani.bits-pilani.ac.in} 
Souradeep Bhattacharya,$^2$
Kaushar Vaidya,$^1$
and Manan Agarwal$^3$\\
$^{1}$ Department of physics, Birla Institute of Technology and Science-Pilani, 333031 Rajasthan, India\\
$^{2}$ Inter University Centre for Astronomy and Astrophysics, Ganeshkhind, Post Bag 4, Pune 411007, India\\
$^{3}$ Department of Physics and Kavli Institute for Astrophysics and Space Research, Massachusetts Institute of Technology, Cambridge, MA 02139, USA
}
\date{Accepted}

\pubyear{2022}

\begin{document}
\label{firstpage}
\pagerange{\pageref{firstpage}--\pageref{lastpage}}
\maketitle
\begin{abstract}
Blue straggler stars (BSS) are peculiar objects which normally appear as a single broad sequence along the extension of the main sequence. Only four globular clusters (GCs) have been observed to have two distinct and parallel BSS sequences. For the first time for any open cluster (OC), we report double BSS sequences in Berkeley 17. Using the machine-learning based membership algorithm ML-MOC on \textit{Gaia} EDR3 data, we identify 627 cluster members, including 21 BSS candidates out to 15$\arcmin$ from the cluster center. Both the BSS sequences are almost equally populated and parallel to one another in Gaia as well as in Pan-STARRS colour-magnitude diagram (CMD). We statistically confirm their presence and report that both BSS sequences are highly segregated compared to the reference population out to $\sim$5.5$\arcmin$ and not segregated thereafter. The lower densities of OCs make BSS formation impossible via the collisional channel. Therefore, mass transfer seems to be the only viable channel for forming candidates of both sequences. The gap between the red and blue BSS sequences, on the other hand, is significant and presents a great opportunity to understand the connection between BSS formation and internal as well as external dynamics of the parent clusters.
\end{abstract}
\section{Introduction}
Blue straggler stars (BSS) are late bloomers, still lingering on the main-sequence, and lagging compared to their other massive siblings which have already evolved off the main-sequence \citep{Sandage1953}. Since their first finding, as observational and computational techniques improved, they became a common exotic stellar population to a variety of stellar environments such as open clusters \citep[OCs;][]{Ahumada2007,Vaidya2020,Rain2021}, globular clusters \citep[GCs;][]{Fusi1992,Sarajedini1993}, dwarf galaxies \citep{Momany2007,Mapelli2009}, and Galactic fields \citep{Preston2000}. Formation channels of BSS are well established. There are three main channels through which BSS can form: (i) stellar collisions – direct collision between single stars \citep{Hills1976} and a stellar collision in dynamical interaction of binaries with single stars or with another binary \citep{Leonard1989,Leigh2019}, (ii) the merger of an inner binary in a hierarchical triple system through the Kozai mechanism \citep{Perets2009,Naoz2014}, and (iii) mass transfer in a primordial binary system \citep{McCrea1964}.

In contrast to this, two equally populated parallel BSS sequences separated by color have been detected in central regions of three core-collapsed GCs, M30 \citep{Ferraro2009}, NGC 362 \citep{Dalessandro2013}, M15 \citep{Beccari2019}, and one GC, NGC 1261 \citep{Simunovic2014} which is of intermediate dynamical age \citep{Raso2020}. These studies as well as several follow-up studies \citep{Xin2015,Portegies2019} demonstrated that the blue BSS sequence form via stellar collisions and the red BSS sequence form via the mass-transfer channel as a consequence of increased density in the core during the core-collapse process. Although \citet{Knigge2009} have shown that mass transfer is the dominant BSS formation channel, as mentioned earlier, both formation channels coexist in the four GCs with almost similar rates of BSS formation.
\begin{figure*}
	\includegraphics[width=0.95\textwidth]{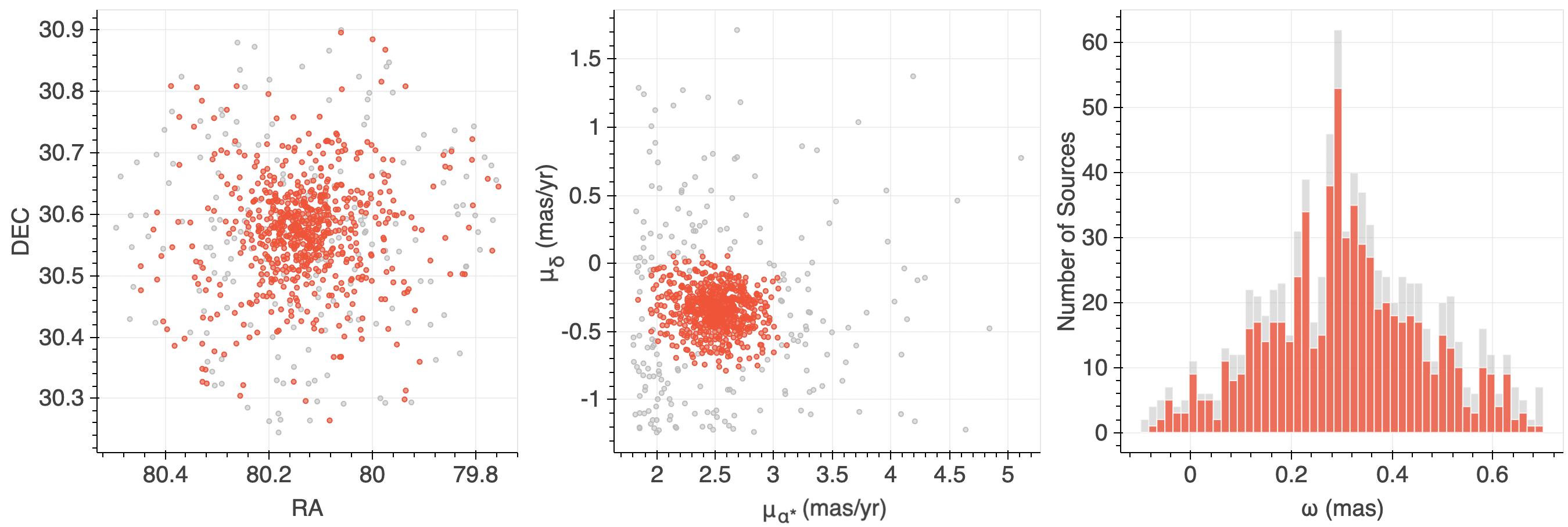}
   	\vspace{-0.2cm}
    \caption{The spatial, proper motion and parallax distribution of \textit{sample} \textit{sources} (gray) and cluster members (orange) identified by the ML-MOC algorithm.}
    \label{fig:memb_figure}
\end{figure*}
In the present work, we report the discovery of two BSS sequences in the Galactic OC Berkeley 17 (RA=05:20:37, DEC=$+$30:35:12, J2000). With an estimated age of 8.5 -- 10 Gyr \citep{Phelps1997,Bragaglia2006}, it is among the oldest OCs yet discovered. It lies along the Galactic anti-direction at a distance of $\sim$3100$^{+285.5}_{-352.9}$ pc \citep{Bhattacharya2019,Cantat2018}. It has a metallicity of [Fe/H] $\sim-$0.1$\pm$0.09 dex \citep{Friel2005}. BSS candidates in Berkeley 17 were reported from photometric data by \citet{Ahumada2007} and later by \citet{Bhattacharya2017}. Using  \textit{Gaia} DR2 \citep{GaiaDR22018} photometric and astrometric data, the list of BSS candidates in Berkeley 17 were refined by \citet{Bhattacharya2019} and \citet{Rain2021}.

The rest of the paper is organised as follows. In \S\ref{sec:data}, we explain the membership identification process of the cluster from Gaia Early Data Release 3 \citep[\textit{Gaia} EDR3;][]{Gaiaedr32021} data. In \S\ref{sec:results}, we perform reddening correction to the cluster members, identify BSS population of the cluster, and statistically confirm the presence of the double BSS sequences. In \S\ref{sec:discussion}, we explore the various possibilities for the presence of the double sequences and analyse the findings. Finally, in \S\ref{sec:summary}, we summarize the current work.
\begin{figure*}
    \centering
	\begin{subfigure}[b]{0.32\textwidth}
    		\includegraphics[width=0.98\textwidth]{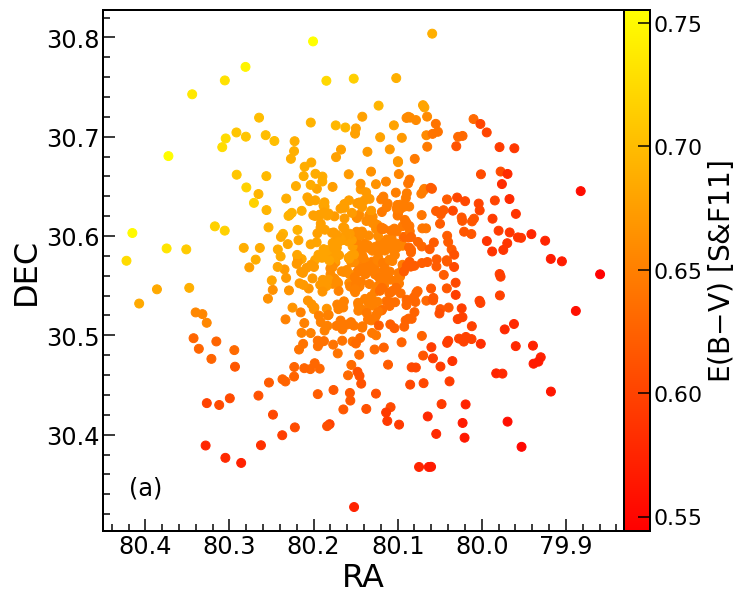}
		\caption*{}
	\end{subfigure}
	\quad
	\begin{subfigure}[b]{0.32\textwidth}
   		\includegraphics[width=0.9\textwidth]{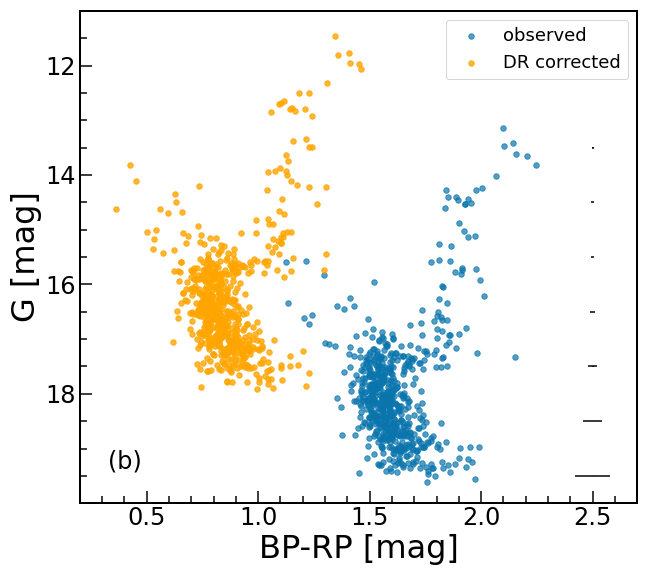}
		\caption*{}
	\end{subfigure}
	\hspace{-0.34cm}
	\begin{subfigure}[b]{0.32\textwidth}
   		\includegraphics[width=0.9\textwidth]{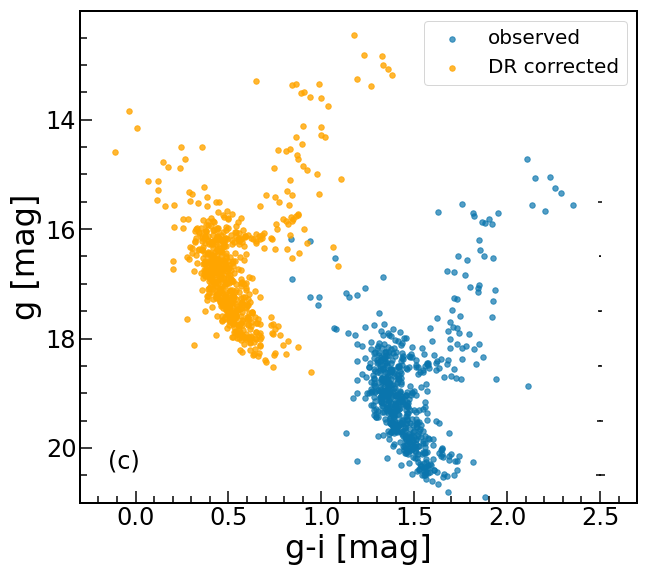}
		\caption*{}
	\end{subfigure}
	\vspace{-0.6cm}
    \caption{(a) The S$\&$F11 reddening map of the cluster members. (b) The \textit{Gaia} EDR3 observed (blue) and differential reddening and extinction corrected CMDs (orange). (c) The PS1 DR1 observed (blue) and differential reddening and extinction corrected CMDs (orange). The errors in colors with respect to magnitudes of \textit{Gaia} EDR3 and PS1 DR1 CMDs are shown as horizontal dashes.}
    \label{fig:DR_correction}
\end{figure*}
\section{Membership identification}
\label{sec:data}
In order to identify the cluster members, we implement the ML-MOC algorithm \citep{Agarwal2021} on the \textit{Gaia} EDR3. We briefly discuss the method for identifying cluster members here (see \citealt{Agarwal2021} for detailed information on the ML-MOC). ML-MOC is a machine-learning based algorithm that uses the k-Nearest Neighbour \citep[kNN;][]{Cover_kNN} and the Gaussian mixture model \citep[GMM;][]{mclachlan200001} algorithms and does not require any prior information of a cluster. First, we take all \textit{Gaia} EDR3 sources within 20$\arcmin$ radius from the cluster center that are having non-zero astrometric and photometric parameters, as well as errors in G mag smaller than 0.005, to create \textit{All} \textit{Sources}. Then kNN is used to exclude most of the likely field stars and obtain \textit{sample} \textit{sources} based only on proper motion and parallax information. Then GMM is applied to proper motions and parallaxes to separate cluster members from noise while giving a membership probability. This is extensively tested in \citet{Agarwal2021} for \textit{Gaia} DR2 and \citet{Bhattacharya2022} for \textit{Gaia} EDR3. 

We have identified 669 sources as cluster members. The spatial, proper motion, and parallax distributions of \textit{sample} \textit{sources} and the identified cluster members are depicted in Figure \ref{fig:memb_figure}. Comparing with Pan-STARRS DR2 (PS1 DR2) for a representative OC, \citet{Bhattacharya2022} find that \textit{Gaia} EDR3 is 90$\%$ complete down to G$\sim$20 mag. Comparing with spectroscopically identified members of another OC, they find that ML-MOC members have a contamination fraction of 2.3$\%$ down to G = 19.5 mag. The main-sequence turnoff of Berkeley 17 is at G $\sim$18 mag, whereas all the BSS candidates are brighter than G = 17.5 mag; hence, our BSS sample is likely to be complete and least contaminated.

We use cluster members out to 15$\arcmin$, beyond which probable cluster members are indistinguishable from field stars. Figure \ref{fig:DR_correction}(b) shows the BP$-$RP vs G CMD (blue) of the 627 cluster members out to 15$\arcmin$ from the cluster center. We also check the CMD of the cluster members in another photometric system, PS1 DR1 \citep{Flewelling2020}, which are cross-matched to the identified cluster members within 1$\arcsec$ search radius. Figure \ref{fig:DR_correction}(b) shows g$-$i vs g CMD (blue) of the identified Berkeley 17 members.
\vspace{-0.5cm}
\section{Results}
\label{sec:results}
\begin{figure}
    \centering
   		\includegraphics[width=0.45\textwidth]{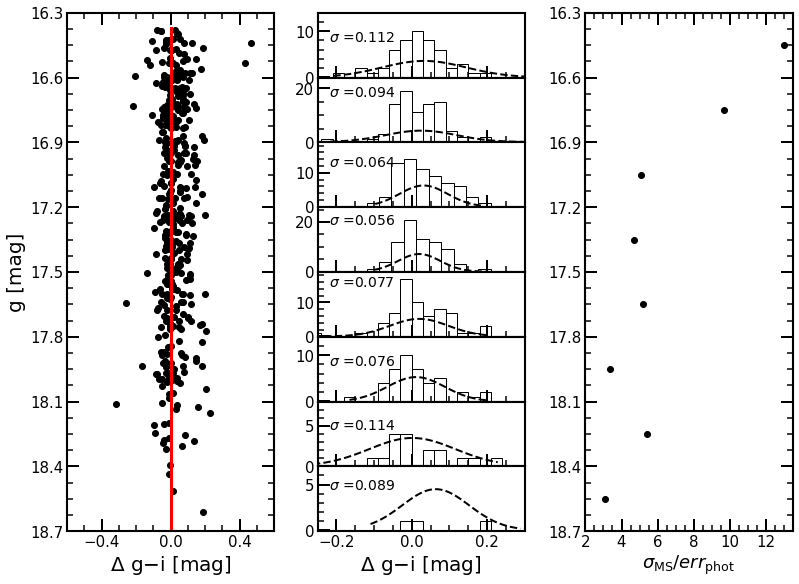}
   		\vspace{-0.13cm}
	\caption{The left panel represent g$-$i mag of main-sequence stars of PS1 DR1 CMD normalized with respect to the plotted PARSEC isochrone. The plots in middle panel represent distribution of normalized g$-$i mag in bins of 0.3 magnitude width fitted with Gaussian distributions, where values of $\sigma$ show spread of g$-$i distributions. In the right panel, the ratio of reddening corrected main-sequence widths to photometric errors are plotted with respect to magnitudes.}
	\label{fig:MS_spread_vs_phot_error}
\end{figure}
\begin{figure*}
    \centering
	\begin{subfigure}[b]{0.48\textwidth}
    		\includegraphics[width=0.9\textwidth]{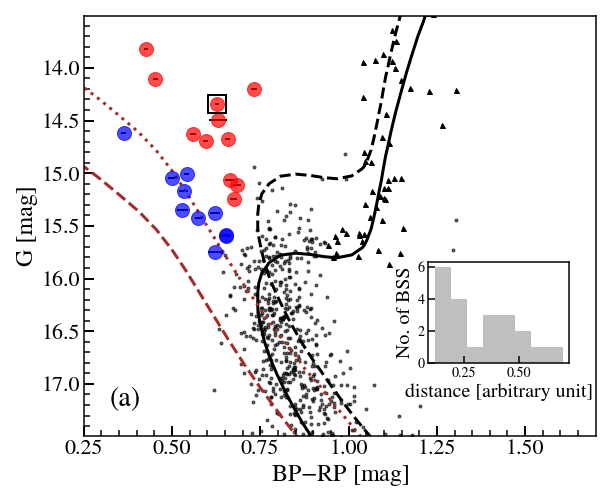}
		\caption*{}
	\end{subfigure}
	\begin{subfigure}[b]{0.48\textwidth}
   		\includegraphics[width=0.9\textwidth]{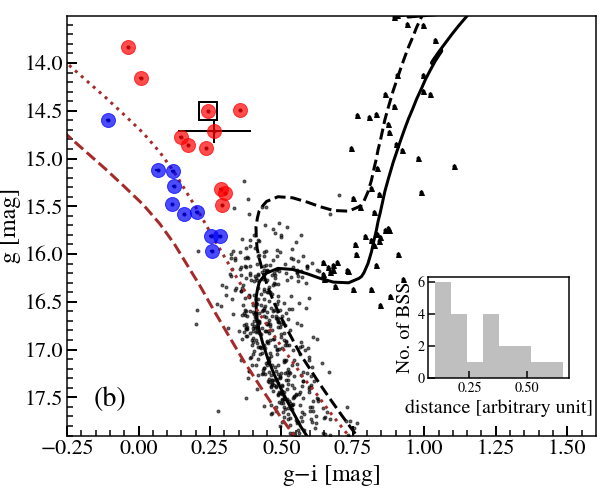}
		\caption*{}
	\end{subfigure}
	\vspace{-0.9cm}
    \caption{The \textit{Gaia} EDR3 CMD (a) and PS1 DR1 CMD (b) of the reddening corrected members. The red and blue BSS candidates are shown as blue and red filled circles respectively with black error bars, the cluster members are shown as black dots, and the reference population (REF) is shown as little triangles. The black open square shows the BSS candidate having RUWE = 3.53. The black solid line shows the plotted PARSEC isochrone, the black dashed line shows the PARSEC isochrone shifted up by 0.75 mag to locate equal mass binaries, the brown dashed line shows ZAMS set at an age of 160 Myr, and the brown dotted line shows ZAMS shifted up by 0.75 to locate equal mass binaries. The inset plots represent histograms of distances of BSS candidates from the ZAMS.}
    \label{fig:cmds}
\end{figure*}
\vspace{-0.1cm}
\subsection{Reddening correction}
In order to correct each cluster member from interstellar dust present along the line of sight of the cluster, we use the all-sky  dustmap provided by \citet[hereafter S$\&$F11]{Schlafly2011}. We convert the extinction values of each star into extinctions in G, BP, and RP filters of \textit{Gaia} EDR3, and g and i filters of PS1 DR1 using equations 1, 3a and 3b of \citet{Cardelli1989}. Figure \ref{fig:DR_correction}(a) shows the reddening map of Berkeley 17. The reddening variation along the line of sight of the cluster is $\sim$0.25 mag. Figure \ref{fig:DR_correction}(b) and \ref{fig:DR_correction}(c) show the observed (blue) and differential reddening corrected (orange) CMDs in the \textit{Gaia} EDR3 and PS1 DR1 photomeric systems, with their photometric errors. 

The main-sequences of both \textit{Gaia} EDR3 and PS1 CMDs remain broad even after reddening correction. As can be seen from Figure \ref{fig:DR_correction}(c) the photometric errors in PS1 DR1 are quite smaller compared to \textit{Gaia} EDR3, therefore, we check for PS1 DR1 CMD that how the main-sequence width vary in magnitude with respect to its photometric errors. To estimate spread in the main-sequence after the reddening correction, we have normalized main-sequence of the corrected CMD with respect to plotted isochrone, as shown in the left panel of Figure \ref{fig:MS_spread_vs_phot_error}. We then divided main-sequence into 8 bins of G = 0.3 mag width and plotted distributions of g$-$i of each bin and fitted with Gaussian distributions. The corrected PS1 DR1 CMD is shifted up by $\sim$2.0 mag compared to its observed CMD. Therefore, we select 8 bins of 0.3 mag width from the observed PS1 DR1 CMD, whose magnitude ranges are the same as the bins chosen to estimate main-sequence width after accounting for the 2.0 magnitude shift. We then plotted the ratio of main-sequence widths to photometric errors with respect to magnitudes as shown in the right panel Figure \ref{fig:MS_spread_vs_phot_error}. From the right panel of Figure \ref{fig:MS_spread_vs_phot_error}, we can see that the $\sigma_{\text{MS}}$/$err_{\text{phot}}$ value is highest at g = 16.3 mag, which is close to the turnoff magnitude of the cluster, and even at fainter magnitudes where $err_{\text{phot}}$ is higher, $\sigma_{\text{MS}}$/$err_{\text{phot}}$ is still $>$ 3. This implies that the main-sequence is quite wide compared to the corresponding photometric errors, which could be due to the combined effect of uncorrected differential reddening and the presence of unresolved binaries. Thus, it shows that there still be uncorrected differential reddening that needs to be corrected for.
\vspace{-0.3cm}
\subsection{The BSS population}
In order to identify BSS candidates of this cluster, we use the BSS classification method adopted by \citet{Rao2021} that focuses on retrieving the most likely BSS candidates while leaving behind sources near the main-sequence turnoff as probable BSS. We have plotted a PARSEC isochrone of age = 8.5 Gyr, [M/H] = $-$0.19 dex, and distance = 2900 pc to the reddening corrected CMDs as shown in Figure \ref{fig:cmds}. We have identified 21 BSS candidates in this cluster.

\citet{Rain2021} identified 20 BSS candidates using the cluster members identified by \citet{Cantat2018}. We have 17 BSS candidates in common with them. Out of their 3 remaining BSS candidates, 1 is near to the main-sequence turnoff of the cluster therefore we do not consider it as a BSS candidate and 2 are not our members. Of our 4 additional BSS candidates, 2 are out of the field of view of the cluster extent used by \citet{Cantat2018} to identify the cluster members and 2 are not BSS candidates according to \citet{Rain2021}. \citet{Bhattacharya2019} identified 14 BSS candidates in this cluster. We have 6 BSS candidates common with them, whereas, 1 of their BSS candidate is our cluster member but not a BSS candidate according to the BSS classification criteria of \citet{Rao2021}, and remaining 7 are outliers in proper motion distribution. Of our 15 additional BSS candidates, 1 is not a BSS candidate according to \citet{Bhattacharya2019} and 14 were identified as non-members by \citet{Bhattacharya2019} because of relatively larger parallax errors in the Gaia DR2 coupled with their overly stringent parallax range chosen for membership determination. 
\vspace{-0.3cm}
\subsection{Identifying the double BSS sequences}
\label{sec:double}
The double BSS sequences are clearly evident in the observed \textit{Gaia} EDR3 and PS1 DR1 CMDs shown in Figure \ref{fig:DR_correction}(b) and \ref{fig:DR_correction}(c), respectively. Despite the reddening correction, the double BSS sequences remain distinct, however, the space between them is slightly reduced compared to the observed CMDs. Figure \ref{fig:cmds}(a) and \ref{fig:cmds}(b) show reddening corrected \textit{Gaia} EDR3 and PS1 DR1 CMDs, with the red and blue BSS sequences represented as red and blue filled circles with black error bars, respectively.

As can be seen in Figure \ref{fig:cmds}, the blue BSS sequence is relatively narrow, whereas the red sequence is more spread out. However, both the sequences are parallel and almost equally populated, consisting of 10 candidates in the blue sequence and 11 candidates in the red sequence. We statistically check the validity of these two sequences from their order of separation in the CMDs in the following manner. Given the small number of BSS, Hartigan’s dip test \citep{Hartigan1985} is not applicable here to show the bimodality's significance. This is why, we follow the indirect approach of checking bimodality using multi-Gaussian fits. We calculate the perpendicular distance of each BSS candidate from the fitted ZAMS and plot a histogram of the estimated distances as shown in the inset plots of Figure \ref{fig:cmds}. We then fit single and double Gaussian distributions to the histograms of the estimated distances in both \textit{Gaia} EDR3 and PS1 DR1 CMDs and estimate Bayesian Information Criteria \citep[BIC;][]{Schwarz1978} for each of the fitted Gaussian distributions. For \textit{Gaia} EDR3 CMD, we get BIC as $-$11.31 and $-$11.32 for the fitted single and double Gaussian distributions, respectively. For PS1 DR1 CMD, we get BIC as $-$11.84 and $-$12.84 for the fitted single and double Gaussian distributions, respectively. From this exercise, we find that BIC is minimum for double Gaussian models for both the CMDs, which shows that the double Gaussian model is the best-fitted model to the histograms of the estimated distances. Hence, it proves the bimodality in the estimated distances. Thereby, it confirms the presence of the double BSS sequences in Berkeley 17 as previously identified only in four GCs M30, M15, NGC 362, and NGC 1261.
\vspace{-0.3cm}
\subsection{Segregation of BSS}
\label{sec:MST}
\begin{figure}
    \centering
   		\includegraphics[width=0.46\textwidth]{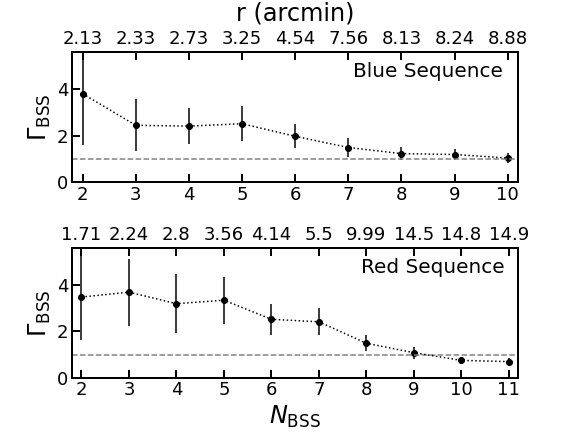}
   		\vspace{-0.15cm}
	\caption{The $\Gamma_{\text{BSS}}$ vs N$_{\text{BSS}}$ profile for the blue BSS sequence and the red BSS sequence. On the top x-axis of both panels, the distances of the BSS candidates from the cluster centre are shown. The error bars show the standard deviation in $\Gamma_{\text{BSS}}$ values. The gray line shows $\Gamma_{\text{BSS}}$ = 1, i.e., no segregation of BSS compared to the REF.}
	\label{fig:crd}
\end{figure}
The red BSS have been found to be more mass segregated than blue ones in the four GCs \citep{Ferraro2009,Dalessandro2013,Simunovic2014,Beccari2019}. Hence, we check for the mass segregation of the two BSS sequences in Berkeley 17. The Anderson-Darling test \citep{Stephens1974} and Kolmogorov-Smirnov test \citep{Chakravarti1967} were used to check if BSS and the reference population (REF) were drawn from the different distributions, but with p-value $>$ 0.05 for both tests, the results were inconclusive (likely because these tests are weak for low number statistics, \citealt{Razali2011}). We employ the minimum spanning tree (MST) method in order to show the level of segregation of both BSS sequences in comparison to REF. We use sub giant branch stars, red giant branch stars, and red clump stars as REF, which are depicted as small triangles in Figure \ref{fig:cmds}. The MST of a sample is the shortest available path length that connects all its data points without forming a closed loop. The more compact or segregated data would have a lower value of MST length. This method has been widely used to detect mass segregation in OCs \citep{Allison2009}. 

We select N BSS from one of the BSS sequences and the same number of random sources from REF. We then estimate the degree of segregation of BSS candidates ($\Gamma_{\text{BSS}}$) from the ratio of MST lengths of REF and BSS. The errors associated with $\Gamma_{\text{BSS}}$ are estimated by taking the standard deviation of 100 values of $\Gamma_{\text{BSS}}$ obtained by iterating the same process for 100 times \citep{Tarricq2022,Bhattacharya2022}. Figure \ref{fig:crd} shows the $\Gamma_{\text{BSS}}$ vs N$_{\text{BSS}}$ profiles for the blue and the red BSS sequences. The gray line at $\Gamma_{\text{BSS}}$ = 1 depicts no BSS segregation, indicating that the MST lengths of BSS candidates and REF are equal. For the blue sequence, we find $\Gamma_{\text{BSS}}$ > 1 within errors up to N$_{\text{BSS}}$ = 6 which is at a distance of 4.54$\arcmin$ from the cluster center. For the red sequence, we find $\Gamma_{\text{BSS}}$ > 1 within errors up to N$_{\text{BSS}}$ = 7 which is at a distance of 5.5$\arcmin$ from the cluster center. Thus, we infer that both the BSS sequences are segregated out to  $\sim$5.5$\arcmin$, after that segregation has not been observed in any of the BSS sequence. The red sequence ($\Gamma_{\text{BSS}}$ = 2.41) appears more segregated than the blue sequence ($\Gamma_{\text{BSS}}$ = 1.97) at the same radial range but the values overlap within 1 sigma uncertainty. Our findings are consistent with \citet{Bhattacharya2019} and \citet{Rao2021}. They reported that Berkeley 17 is in a stage of intermediate dynamical evolution where BSS of only the inner-most region of the cluster are segregated.
\vspace{-0.5cm}
\section{Discussion}
\label{sec:discussion}
The blue BSS sequences identified in the four GCs \citep{Ferraro2009,Dalessandro2013,Simunovic2014,Beccari2019} were quite narrow and nicely correlated with collisional isochrones \citep{Sills2009}, implying the formation of all blue BSS around the same time via stellar collisions. On the other hand, the red BSS sequences were comparatively scattered, and $\sim$0.75 mag brighter than the zero-age main sequence (ZAMS), which approximately represents the low luminosity boundary formed by binaries with ongoing mass transfer \citep{Tian2006}, implying formation via the mass-transfer channel. Since the blue BSS sequence of Berkeley 17 is narrow and follows a straight line, it appears that all its BSS candidates have been formed almost around the same time, whereas red BSS sequence of Berkeley 17 is rather dispersed.

The BSS sequences detected in Berkeley 17 are similar to the double sequences of BSS earlier identified in the four GCs in terms of the photometric distribution but not in terms of the spatial distribution. The BSS candidates of both the sequences in the four GCs are located only inside the core region, which is not the case for Berkeley 17, both the sequences are observed up to 15$\arcmin$ from the cluster center. The blue BSS candidates are located in the varying cluster density as they span a range of 1.30$\arcmin$ -- 8.9$\arcmin$ in distances from the clusters center. Moreover, OCs are comparatively less dense systems and have not been core-collapsed yet; thus, the likelihood of BSS formation via stellar collisions is negligible. Furthermore, the blue sequence is nicely replicated by the equal mass binary isochrone of ZAMS. The sources on the equal mass binary isochrone of ZAMS are predicted to be in the active phase of mass transfer \citep{Tian2006}. With metallicity of $-$0.19 dex and the amount of reddening correction done, the equal mass binary isochrone of ZAMS perfectly traces the blue sequence. However, since the known metallicity of Berkeley 17 ranges from $-$0.19 dex to $-$0.01 dex \citep{Friel2005}, a change in metallicity can shift the equal mass binary isochrone of ZAMS towards the red BSS sequence. It has been revealed through spectroscopic and UV observations that in many OCs, such as NGC 188 \citep{Geller2008,Gosnell2015}, M67 \citep{Geller2015,Sindhu2019,Jadhav2019}, NGC 6819 \citep{Milliman2015}, and NGC 7789 \citep{Nine2020,Vaidya2022}, $\sim$33$\%$ -- 85$\%$ of BSS formed via binary system evolution, primarily through the mass-transfer channel. The triple-mediated interactions are also considered as an important formation channel for some BSS of OCs, such as M67 and NGC 188 \citep{Leigh2011,Leiner2016,Bertelli2018}, Melotte 66 \citep{Rao2022}, and NGC 2506 \citep{Panthi2022}. Furthermore, \citet{Ferraro2009} and \citet{Dalessandro2013} found that blue sequences of M30 and NGC 1261 contain W UMa binaries. Additionally, several studies such as \citet[and references therein]{Jiang2017} have shown that BSS formed via the mass-transfer channel do exist below the low luminosity boundary. Thus, the blue sequence can also have some BSS formed via the mass-transfer channel.

Of the 11 BSS candidates of the red BSS candidates, 7 are centrally concentrated as shown in \S \ref{sec:MST}. Some of them might be the evolved BSS which have been segregated in the cluster center due to the dynamical friction. It has been shown that the red sequence form due to the combination of evolved BSS and primordial binary evolution BSS \citep{Jiang2017}. \citet{Bhattacharya2017} showed that the cluster is mass segregated and has a core-tail morphology using the cluster members identified from PS1 DR1 data. \citet{Rao2021} also inferred that the cluster is of intermediate dynamical age based on its A$^+_{\mathrm{rh}}$ (the area enclosed between cumulative radial distributions of BSS and reference population up to half-mass radius from the cluster center). These previous studies show that Berkeley 17 is dynamically evolved enough where mass segregation is taking place, therefore it is very likely that some of the BSS candidates of red sequence are evolved BSS candidates segregated in the cluster center due to the dynamical friction. One red BSS candidate (Gaia EDR3$\_$ID = 3446810558983275904) that is located at $\sim$14.8$\arcmin$ from the cluster center, shown as black open square in Figure \ref{fig:cmds}, has renormalised unit weight error (RUWE) as 3.53 in \textit{Gaia} EDR3 data. In \textit{Gaia} DR2 data, it has RUWE = 3.31. \citet{Penoyre2022} prescribed that if RUWE is consistently increasing from \textit{Gaia} DR2 to \textit{Gaia} EDR3, then the source is likely to have second gravitating companion. The RUWE value is indeed increasing in the two epochs of data. Thus mass-transfer is the likely origin for this BSS candidate.

Until now, BSS have been observed as one broad sequence above the turnoff of main sequence of OCs and GCs, except for the four GCs. Thus, double BSS sequences in OCs have not been observed, although they are present in Berkeley 17. As explained previously, in OCs, BSS form primarily through the mass-transfer channel. Furthermore, some of the BSS generated by the mass-transfer channel in M30 were identified below the low-luminosity boundary. Similarly, if the BSS of both the sequences of Berkeley 17 are formed via the mass-transfer channel, then the key question here is, why is there such a large gap between these two BSS sequences? We speculate on the three potential reasons behind the formation of the two sequences as described below.
\begin{enumerate}
    \item \textit{There is a significant difference between rotational velocities of BSS candidates of the two sequences.} The projected rotational velocities of BSS of two OCs, NGC 188 and NGC 6819, were found to vary substantially from \textit{v} sin \textit{i} $\leq$ 10 km/s to \textit{v} sin \textit{i} = 50 km/s \citep{Mathieu2009}. On the other hand, \citet{Lovisi2013} performed spectroscopic observations of 15 BSS candidates, 7 along the blue sequence and 8 along the red one of M30 GC, using FLAMES@VLT and XSHOOTER@VLT. They demonstrated that all 15 BSS candidates have projected rotational velocities as $\sim$30 km/s, with the exception of one blue BSS candidate that is a fast rotator (\textit{v} sin \textit{i} > 90 km/s) and a W UMa binary. It is still possible that the two BSS sequences consist of two separate populations of slow and fast rotators. Therefore, one needs to investigate this possibility in detail. 
    \item \textit{There is a presence of multiple stellar populations (MPs) in this old OC.} MPs are the occurrence of star-to-star variations in chemical abundances of a cluster that are not predicted from stellar evolutionary processes but are only feasible owing to the formation of cluster members over a period of time or in multiple episodes of star formation \citep{Bastian2018}. Since Berkeley 17 is among the oldest OCs, it might have swallowed a younger cluster during its lifetime. Indeed \citet{Piatti2022} have observed such a collision of OCs. If such a collision with a younger cluster happened to Berkeley 17 in its past, then the Blue BSS sequence might simply be younger stars from this swallowed cluster.
    \item \textit{There is uncorrected differential reddening.} As shown in Figure \ref{fig:MS_spread_vs_phot_error}, the spread in main-sequence is quite higher compared to the photometric errors which could be due the uncorrected reddening and presence of unresolved binaries. If the main-sequence spread is due to uncorrected differential reddening, this may also be affecting the BSS positions in the CMD, giving the appearance of two sequences. The S$\&$F11 map used to correct the cluster members from extinction and reddening is of low resolution\footnote{We attempted DR correction with the relatively higher resolution Bayestar map \citep{Green2019}, but that fared worse than the S$\&$F11 map used here, giving an even broader main-sequence.}. Future high-resolution dustmaps will help us to check if differential reddening is the possible cause for the presence of double BSS sequences in Berkeley 17.
    \item If three former assumptions are invalid, then it is merely a coincidence that the combination of two or more of factors discussed above projected BSS population on the CMD in such a way that two sequences are observed.
\end{enumerate}
High-resolution spectroscopic observations are required in order to investigate all three speculations made here and unveil the reason behind the double sequences in Berkeley 17. The measurement of radial velocities will first help in securing the membership of all the BSS. The multi-epoch radial velocity information will help to discover the binaries and confirm the formation mechanism. The mass-transfer from an evolved donor, such as asymptotic giant branch stars, can considerably contaminate the surface of BSS with s-process elements, barium, carbon, oxygen, and nitrogen \citep{Sivarani2004,Milliman2015}. If BSS are severely deficient in carbon and oxygen as compared to main-sequence stars, a red giant branch star would be the progenitor of BSS formation \citep{Ferraro2006}. BSS formed via collisions, mergers, or mass-transfer from a main-sequence star, on the other hand, exhibit no abundance variation when compared to main-sequence stars. Thus, the high-resolution spectra will also help in determining the abundances of BSS, which will help in pinpointing their formation mechanisms.
\section{Summary}
\label{sec:summary}
For the first time, we have detected two parallel and almost equally populated BSS sequences in the Galactic OC, Berkeley 17. Though double BSS sequences have only been found in four GCs and are unprecedented in OCs, they are clearly visible in Berkeley 17. It has been shown that in the four GCs, one sequence forms via the collisional channel and another via the mass-transfer channel. The collisional channel is unfavorable for OCs owing to their lower density. If both the sequences in Berkeley 17 have been formed via the mass-transfer channel, the enormous gap between the two sequences is unusual. We speculate on a few possible explanations for their origin, such as differences in the rotational velocities, the existence of multiple stellar populations, or a mere coincidence. However, future high-resolution spectroscopic observations may shed light on the reason for their presence.
\vspace{-0.5cm}
\section*{Acknowledgements}
We thank the anonymous referee and Dr. Giacomo Beccari (European Southern Observatory, Garching, Germany) for their valuable comments which improved the quality of the paper. SB is funded by the INSPIRE Faculty award (DST/INSPIRE/04/2020/002224), Department of Science and Technology (DST), Government of India. This work has made use of early third data release from the European Space Agency (ESA) mission {\it Gaia} (\url{https://www.cosmos.esa.int/gaia}), Gaia eDR3 \citep{Gaiaedr32021}, processed by the {\it Gaia} Data Processing and Analysis Consortium (DPAC, \url{https://www.cosmos.esa.int/web/gaia/dpac/consortium}). This research has made use of the VizieR catalog access tool, CDS, Strasbourg, France, Astrophysics Data System (ADS) governed by NASA (\url{https://ui.adsabs.harvard.edu}), {\small ASTROPY}, a {\small PYTHON} package for astronomy \citep{Astropy2013}, {\small NUMPY} \citep{Harris2020}, {\small MATPLOTLIB} \citep{Hunter4160265}, and {\small SCIPY} \citep{Virtanen2020}.
\vspace{-0.5cm}
\section*{Data Availability}
The data underlying this article are publicly available at \url{https://gea.esac.esa.int/archive}. The derived data generated in this research will be shared on reasonable request to the corresponding author.
\vspace{-0.5cm}
\bibliographystyle{mnras}
\bibliography{References} 
\bsp	
\label{lastpage}
\end{document}